\documentclass[12pt]{iopart}

\usepackage{graphicx}

\newcommand{\eps }{\varepsilon }

\begin{document}
\title[Many-body theory of positron annihilation spectra]
{Many-body theory of gamma spectra from positron-atom
annihilation}
\author{L J M Dunlop and G F Gribakin}

\address{Department of Applied Mathematics and Theoretical Physics, Queen's 
University, Belfast BT7 1NN, Northern Ireland, UK}

\ead{l.dunlop@qub.ac.uk and g.gribakin@qub.ac.uk}
\begin{abstract}
A many-body theory approach to the calculation of gamma spectra of
positron annihilation on many-electron atoms is developed. We evaluate
the first-order correlation correction to the annihilation vertex
and perform numerical calculations for the noble gas atoms.
Extrapolation with respect to the maximal orbital momentum of the
intermediate electron and positron states is used to achieve convergence.
The inclusion of correlation corrections improves agreement with
experimental gamma spectra.
\end{abstract}

\submitto{\JPB}

\maketitle

\section{Introduction}

The main aim of our paper is to develop a many-body theory approach to the
calculation of gamma spectra from positron-atom annihilation.

Positron annihilation has been an important tool for studying electronic
and atomic structure of solids for over half-a-century. Its history can be
traced from the first review of the theory of positron annihilation in solids
by Ferrel (1956) to a more recent work by Puska and Nieminen (1994).
Positron lifetimes and spectra of annihilation gamma quanta contain information
about the structure and composition of bulk materials and surfaces,
presence and concentration of defects and voids or pores, and their sizes,
and electron momentum distribution. Interpretation of positron annihilation
data requires good theoretical understanding of the process. Many-body theory
was the first method used to determine positron annihilation rates in metals
(Kahana 1963, Carbotte 1967). In particular, it was successful in explaining
large enhancement factors that increase the annihilation rates above that
obtained in the noninteracting electron gas approximation (see also Arponen
and Pajanne 1979).

In the 1990's measurements of gamma-ray spectra from positron annihilation
on atoms and molecules in the gas phase became possible (Tang \etal 1992,
Coleman \etal 1994).
For He the experiment showed excellent agreement with the calculated spectrum
obtained from an elaborate variational positron-He wavefunction
(Van Reeth \etal 1996). This work also revealed sizeable deviations of the
shape spectrum from a Gaussian, which is often used as an approximation.
For noble gas atoms of Ar, Kr and Xe, a careful study of the shape of the
511-keV gamma-ray line provided an estimate of the contribution of positron
annihilation with inner-shell electrons (Iwata \etal 1997a).
A large systematic study was conducted for a variety of inorganic molecules,
alkanes, alkenes, aromatics and perfluorinated and partially fluorinated
hydrocarbons (Iwata \etal 1997b). In particular, this work determined
the relative probability of annihilation on the fluorine atoms and on the
C--H bonds for partially fluorinated hydrocarbons. Such information is
important for achieving better understanding of very large positron
annihilation rates on polyatomic molecules (see, e.g., Surko \etal 2005).

On the theory side, much progress in understanding the interaction of
low-energy positrons with many-electron atoms has been obtained by using
many-body theory. Its application to positron-atom collisions was pioneered
by Amusia \etal (1976) who considered positron scattering from He.
Subsequently, the effect of virtual positronium (Ps) formation on
positron-atom interaction was investigated for heavier noble gas atoms
(Dzuba \etal 1993, 1996). These works showed that virtual Ps formation
gives a large contribution to the positron-atom
attraction\footnote{This explained the success of earlier polarized-orbital
calculations in describing positron elastic scattering and annihilation
for noble gas atoms (McEachran \etal 1980 and references
therein). In that approximation the positron was treated as a heavy particle,
and the strength of the positron-atom polarization potential was
overestimated, making up for the complete neglect of virtual Ps
formation.}. It leads to the emergence of positron-atom
virtual and bound states (Dzuba \etal 1995, 1996), and completely alters
the picture of low-energy positron-atom scattering (Gribakin and King 1996).
These papers employed an approximate method of accounting for virtual
Ps formation. Recently, a consistent many-body theory approach has been
developed. It is based on the summation of the electron-positron ladder
diagram series
(Ludlow 2003, Gribakin and Ludlow 2004). Its contribution is especially
prominent in the calculation of the annihilation rate. These and earlier
calculations (Dzuba \etal 1993, 1996) have demonstrated that
electron-positron correlation effects can enhance the positron-atom
annihilation rate as much as $10^3$ times.

In the paper by Iwata \etal (1997a), the fraction of annihilation with
inner-shell electrons was derived by fitting the experimental data with a
linear combination of the gamma spectra for the valence and inner shells.
These spectra were calculated in the simplest approximation, by using the
positron wavefunction in the static field of the atom. In spite
of a complete neglect of electron-positron correlations, the shapes of the
measured gamma spectra were described reasonably well. On the other hand,
some discrepancies between the theoretical and experimental spectra were
obvious, adding uncertainty
to the estimates of the inner-shell annihilation fractions. For example,
the data for Ar was compatible with a zero contribution of
annihilation with $2s$ and $2p$ electrons.

Since correlations effects play such a large role in positron-atom
interactions, they must be included in the proper theory of annihilation
gamma spectra. In what follows we briefly recount the main facts and
formulae concerning positron annihilation and gamma spectra using the
formalism of creation-destruction operators (section \ref{sec:spectr}).
We then proceed to derive the many-body diagrammatic expansion of the
annihilation amplitude, and consider the significance of various terms and
their relation to the total annihilation rate (section \ref{sec:mb}). In
section \ref{sec:diag}, expressions for the 0th and 1st-order
contributions are reduced to products of radial matrix elements, which
can be evaluated numerically. Section \ref{sec:res} reports the results
of such calculations for the noble gas atoms, which confirm that the
correlation corrections to the annihilation amplitude have a marked effect
on the gamma spectra. We also demonstrate the importance of extrapolation
over the maximal orbital angular momentum of the intermediate electron
and positron states. In the Appendix, expressions for the annihilation
amplitudes involving many-particle wavefunctions in coordinate space
are given.

\section{Basic Theory}
\label{sec:spectr}

\subsection{Annihilation operator}\label{subsec:oper}

Due to the conservation of momentum, annihilation of an electron-positron
pair must lead to the emission of at least two photons\footnote{When this
process occurs in an external field, e.g., when the electron is bound in an
atom, annihilation into a single quantum is also possible. However, its
probability is small, see, e.g., Johnson \etal 1964.}. In fact, according
to quantum electrodynamics (QED), annihilation into two photons is possible
only if the total spin $S$ of the pair is zero (Berestetskii \etal 1982).
For $S=1$ the annihilation results in the emission of three
photons\footnote{For $S=0$ annihilation into $2,\,4,\dots $ photons is
possible, while for $S=1$ the number of photons must be odd ($3,\,5,\dots $).
In both cases the process with the smallest number of photons dominates.}.

In QED the process of electron-positron annihilation is described by the
2nd or 3rd-order diagram, depending on the number of photons emitted
(Berestetskii \etal 1982). The photons carry away the energy of the
particles, $\sim 2mc^2$, and have large momenta $\sim mc$, where
$m$ is the electron mass, and $c$ is the speed of light. Intermediate
electron/positron states in the QED diagrams have similar momenta.
They are much larger than the typical momenta of valence and inner-shell
atomic electrons (with the exception of $1s$ electrons in heavy atoms with
$Z\sim 100$), or the positron momentum in a typical experiment\footnote{In
positron annihilation with matter, even if the initial positron is fast,
it quickly loses energy due to inelastic ionizing collisions, and most of
the annihilation events involve slow positrons.}. As a result, the
annihilation amplitude is independent of the electron and positron momenta.
Hence, the {\em effective operator} for annihilation into gamma quanta with
the total momentum ${\bf P}$ is proportional to
\begin{equation}\label{eq:annop}
\sum _{{\bf k}_1, {\bf k}_2}\hat a_{{\bf k}_1}\hat b_{{\bf k}_2}
\delta _{{\bf k}_1+{\bf k}_2,{\bf P}}\,,
\end{equation}
where $\hat a_{{\bf k}_1}$ ($\hat b_{{\bf k}_2}$) is the destruction operator
of the electron (positron) with momentum ${\bf k}_1$ (${\bf k}_2$), and the
$\delta $-function ensures momentum conservation,
${\bf P}={\bf k}_1+ {\bf k}_2$ (Ferrell 1956). Using electron and positron
destruction operators $\hat \psi ({\bf r})$ and $\hat \varphi ({\bf r})$ in
the coordinate representation,
\begin{eqnarray}\label{eq:ak}
\hat a_{\bf k}=\frac{1}{\sqrt{V}}\int \rme^{-i{\bf k}\cdot {\bf r}}
\hat \psi ({\bf r})d{\bf r},\\
\label{eq:bk}
\hat b_{\bf k}=\frac{1}{\sqrt{V}}\int \rme^{-i{\bf k}\cdot {\bf r}}
\hat \varphi ({\bf r}) d{\bf r},
\end{eqnarray}
where $V$ is the normalization volume, the annihilation operator
(\ref{eq:annop}) can be re-written as (Ferrell 1956, Chang Lee 1957)
\begin{equation}\label{eq:annop1}
\hat O_a({\bf P})\equiv \int \rme^{-i{\bf P}\cdot {\bf r}}\hat \psi ({\bf r})
\hat \varphi ({\bf r})d{\bf r}.
\end{equation}
This equation shows that the annihilation of a {\em nonrelativistic}
electron-positron pair occurs when the particles are at the same point.
Physically, this can be explained using the uncertainty principle.
Indeed, the spatial separation between the points where the photons are
produced is about $\hbar /mc$. This is much smaller than the Bohr radius
$a_0=\hbar ^2/me^2$, i.e. the typical atomic size, or the de Broglie
wavelength of the particles involved ($c=137$ in atomic units, where
$\hbar =m=|e|=1$).

In equations (\ref{eq:annop}) and (\ref{eq:annop1}) electron and positron
spin indices have been suppressed. The operator can be used in this
form if the total spin of the electron subsystem is zero. In this case
the annihilation rate is equal to that averaged over the positron spin.
The rate of annihilation into photons with the total momentum ${\bf P}$
is given by the squared modulus of the transition amplitude of
operator (\ref{eq:annop1}). The total annihilation rate is obtained by
integration over $\rmd^3P/(2\pi )^3$ and summation over the final states
of the system. The correct absolute magnitude of the 2-photon and 3-photon
annihilation rates can be determined by comparison with the spin-averaged
positronium annihilation rates (Berestetskii \etal 1982). In the first case
the squared amplitude must be multiplied by $\pi r_0^2c$, while in the second
case -- by $[4(\pi^2 -9)/3] r_0^2\alpha c$, where $r_0=e^2/mc^2$ is the
classical electron radius, and $\alpha =e^2/\hbar c$ is the fine structure
constant. This shows that 3-photon annihilation is about $10^3$
times slower than 2-photon annihilation, and can often be neglected.

\subsection{Spectrum of photons and annihilation rates}\label{subsec:spra}

Let $|i\rangle $ be the state of $N$ electrons and the positron before
the annihilation, and $|f\rangle $ -- the state of $N-1$ electrons after
the annihilation. In two-photon annihilation, the momentum distribution of
the photons,
\begin{equation}\label{eq:wP}
W_f({\bf P})=\pi r_0^2c\left|\langle f|\hat O_a({\bf P})|i\rangle \right|^2,
\end{equation}
determines their energy spectrum. The total energy of the two photons is
\begin{equation}\label{eq:Egam}
E_{\gamma 1}+ E_{\gamma 2}= 2mc^2+E_i-E_f, 
\end{equation}
where $E_i$ and $E_f$ are the energies of the initial and final states (not
including the rest energy of the constituent particles), and total photon
momentum is ${\bf p}_{\gamma 1}+{\bf p}_{\gamma 2}={\bf P}$.

For ${\bf P}=0$ the photons are emitted in the opposite direction,
${\bf p}_{\gamma 1}=-{\bf p}_{\gamma 2}$, and have equal energies,
$E_{\gamma 1}=E_{\gamma 2}= mc^2+\frac12 (E_i-E_f)\equiv E_\gamma $.
For ${\bf P}\neq 0$ the photon energy is Doppler shifted, e.g.,
for the first photon $E_{\gamma 1}=E_\gamma +Vmc\cos \theta $, where 
${\bf V}={\bf P}/2m$ is the centre-of-mass velocity of the electron-positron
pair, $\theta $ is the angle between the direction of the photon and
${\bf V}$, and we assume that $V\ll c$ and
$p_{\gamma 1}=E_{\gamma 1}/c\approx mc$. Hence, the shift of the photon
energy from the centre of the line, $\epsilon = E_{\gamma 1}-E_\gamma  $, is
\begin{equation}\label{eq:eps}
\epsilon =\frac{Pc}{2}\cos \theta ,
\end{equation}
and the photon energy spectrum is given by
\begin{equation}\label{eq:weps}
w(\epsilon )=\int W_f({\bf P})\delta \left(\epsilon -\case{1}{2} Pc\cos \theta
\right) \frac{\rmd^3P}{(2\pi )^3}.
\end{equation}
Using polar coordinates and averaging over the direction of emission
of the photon, one obtains:
\begin{equation}\label{eq:wepsP}
w(\epsilon )=\frac{1}{c}\int \int _{2|\epsilon|/c}^{\infty}
W_f({\bf P}) \frac{P\rmd P\rmd\Omega _{\bf P}}{(2\pi )^3}.
\end{equation}

On the other hand, using Cartesian coordinates and choosing the $z$ axis along
the direction of the photon, one obtains from (\ref{eq:weps}):
\begin{equation}\label{eq:wepsC}
w(\epsilon )=\frac{2}{c}\int W_f(P_x,P_y,2\epsilon /c)
\frac{\rmd P_x\rmd P_y}{(2\pi )^3}.
\end{equation}
This form shows that the energy spectrum is proportional to the probability
density for a component of the momentum ${\bf P}$. Equation
(\ref{eq:wepsC}) allows one to link $w(\epsilon )$ to the quantity measured
by the angular correlation of annihilation radiation (ACAR) technique.
In ACAR one measures the small angle
$\Theta $ between the direction of one photon and the plane containing the
other photon\footnote{This is so-called 1D-ACAR. A more advanced technique,
2D-ACAR, involves measuring the angle between the directions of the photons,
i.e. {\em two} projections of the momentum. The corresponding distribution is
proportional to $\int W_f(P_x,P_y,P_z)dP_z$, and is useful for studying
the electron density and momentum distribution in anisotropic systems,
e.g., solids (Puska and Nieminen 1994).}. If the direction
perpendicular to the plane is $x$, then
$\Theta =P_x/mc$. Given that the distributions of $P_x$, $P_y$ and $P_z$ are
identical (in an isotropic system), one obtains the distribution of $\Theta $
from $w(\epsilon )$ by a simple change of variable, $\Theta = 2\epsilon /mc^2$.

When a low-energy positron annihilates with a bound electron whose energy
is $\eps _n$, the centre of the photon spectrum is shifted by 
$\eps _n/2$ relative to $mc^2$. The width of the two-photon momentum
distribution is determined by the typical momenta of the bound electron,
$P\sim (2m|\eps _n|)^{1/2}$. The corresponding Doppler width of the
annihilation spectrum, $\epsilon \sim Pc\sim (|\eps _n|mc^2)^{1/2}$,
is much greater than its shift. Hence, one can regard the line as
centred on $E_\gamma = mc^2=511$ keV, even for the annihilation on
inner-shell electrons.

The total annihilation rate in the state $|i\rangle $ is obtained by
integration over the momentum ${\bf P}$ and summation over the final states,
\begin{eqnarray}\label{eq:lambda}
\lambda &=\sum _f \int W_f({\bf P})\frac{\rmd^3P}{(2\pi )^3}\\
        &=\pi r_0^2c \sum _f \int
\langle i|\hat O_a^\dagger ({\bf P})|f\rangle
\langle f|\hat O_a({\bf P})|i\rangle \frac{\rmd^3P}{(2\pi )^3} .
\label{eq:intsum}
\end{eqnarray}
Using closure in the subspace of $(N-1)$-electron states,
$\sum _f|f \rangle \langle f|= 1$, substituting (\ref{eq:annop1})
and integrating over ${\bf P}$, we have:
\begin{equation}\label{eq:lambda1}
\lambda =\pi r_0^2c \int \langle i|\hat n_-({\bf r})\hat n_+({\bf r})|
i\rangle \rmd{\bf r},
\end{equation}
where $\hat n_-({\bf r})=\hat \psi ^\dagger ({\bf r})\hat \psi ({\bf r})$ and
$\hat n_+({\bf r})=\hat \varphi ^\dagger ({\bf r})\hat \varphi ({\bf r})$
are the electron and positron density operators. The annihilation rate
is thus given by the expectation value of the electron density at the
positron integrated over the positron coordinates.

Equation (\ref{eq:lambda1}) gives the two-photon annihilation
rate for a bound positron-atom state $|i\rangle $. The annihilation
rate for a positive-energy positron moving through a gas of atoms is
given by a similar formula (Fraser 1968),
\begin{equation}\label{eq:Zeff}
\lambda =\pi r_0^2c nZ_{\rm eff},
\end{equation}
where  $n$ is the number density of the gas, and
\begin{equation}\label{eq:Zeff1}
Z_{\rm eff}=\int \langle i|\hat n_-({\bf r})\hat n_+({\bf r})|
i\rangle \rmd{\bf r}.
\end{equation}
Here the state $|i\rangle $ describes a collision between the positron with
momentum ${\bf k}$ and the atom (usually in the ground state). It is
normalised to the positron plane wave $\rme^{i{\bf k}\cdot {\bf r}}$ at large
positron-atom separations, i.e., to one positron per unit volume.
The dimensionless parameter $Z_{\rm eff}$ is interpreted as the {\em effective
number} of target electrons which contribute to the annihilation.

If one neglects the interaction between the positron and the atom and writes
the initial state as $|i\rangle =\sqrt{V}b_{\bf k}^\dagger |0\rangle $,
where $|0\rangle $ is the ground state of the $N$-electron atom,
equation (\ref{eq:Zeff1}) gives $Z_{\rm eff}=N$. In reality,
the values are $Z_{\rm eff}$ can be very different from the actual number
of electrons (see, e.g., Surko \etal 2005). Repulsion from the nucleus
prevents the positron from penetrating into the atom, suppressing the
probability of its annihilation with the inner electrons. On the other hand,
the long-range positron-atom attraction increases the positron density
near the atom, making $Z_{\rm eff}$ larger. The Coulomb attraction
within the annihilating electron-positron pair increases $Z_{\rm eff}$ further
(Dzuba \etal 1996, Gribakin and Ludlow 2004).

Note that the rate (\ref{eq:Zeff}) is related to the positron-atom annihilation
cross section $\sigma _a$ by $\lambda =\sigma _anv$, where $v$ is the incident
positron velocity, so that $\sigma _a=\pi r_0^2 Z_{\rm eff}c/v$. On the other
hand, equation (\ref{eq:Zeff1}) shows that $Z_{\rm eff}$ has the form of
a transition amplitude between two identical states $|i\rangle $.
Hence, a perturbation series expansion for this quantity can be
developed (Dzuba \etal 1993).

\section{Many-body theory}\label{sec:mb}
\subsection{Bases and building blocks}\label{subsec:bb}

\begin{figure}[ht]
\begin{center}
\includegraphics*[width=8cm]{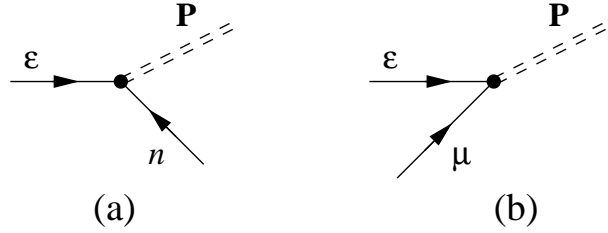}
\end{center}
\caption{Annihilation amplitude for a positron in state $\eps $
and electron in state $n$ occupied in the atomic ground state (a), or an
excited electron in state $\mu $ (b). The double dashed line corresponds to the
annihilation gamma quanta with momentum ${\bf P}$.}\label{fig:ampl}
\end{figure}

When studying positron-atom interactions by many-body theory methods, it
is convenient to use the bases of electron and positron states of the
Hartree-Fock Hamiltonian of the ground-state atom. Denoting the corresponding
electron (positron) wavefunctions and destruction operators by
$\psi _\mu $ ($\varphi _\nu $) and $\hat a_\mu $ ($\hat b_\nu $),
respectively, we have:
\begin{equation}\label{eq:hatpsi}
\hat \psi ({\bf r})=\sum _\mu \psi _\mu ({\bf r})\hat a _\mu ,
\quad
\hat \varphi ({\bf r})=\sum _\nu \varphi _\nu ({\bf r})\hat b_\nu .
\end{equation}
Substituting into (\ref{eq:annop1}) we obtain the effective annihilation
operator in the form,
\begin{equation}\label{eq:annop2}
\hat O_a({\bf P})=\sum _\mu \sum _\nu
\langle {\bf P}|\delta |\mu \nu \rangle \hat a_\mu \hat b_\nu ,
\end{equation}
where
\begin{eqnarray}\label{eq:mael}
\langle {\bf P}|\delta |\mu \nu \rangle &=\int \rme^{-i{\bf P}\cdot {\bf r}}
\psi _\mu ({\bf r})\varphi _\nu ({\bf r}) \rmd{\bf r}\\ \nonumber
&\equiv \int \rme^{-i{\bf P}\cdot ({\bf r}_1+{\bf r}_2)/2}
\delta ({\bf r}_1-{\bf r}_2)
\psi _\mu ({\bf r}_1)\varphi _\nu ({\bf r}_2) \rmd{\bf r}_1 \rmd{\bf r}_2.
\end{eqnarray}
The second form explains the presence of $\delta $ in our notation
for the amplitude. Note that the sum over $\mu $ in
equation (\ref{eq:annop2}) contains two distinct contributions,
$\sum _\mu =\sum _{\mu \leq F} + \sum _{\mu >F}$. The first sum includes
the electron states occupied in the atomic ground state, i.e., those at or
below the Fermi level $F$ (``holes''). The corresponding terms in the
operator (\ref{eq:annop2}) describe positron annihilation which leads to
creation of a hole, as shown by diagram (a) in figure \ref{fig:ampl}. The
second sum is over the electron states above the Fermi level (``particles''),
and describes positron annihilation with an excited electron, diagram (b) 
in figure \ref{fig:ampl}.

Substituting (\ref{eq:hatpsi}) and their analogues for
$\hat \psi ^\dagger ({\bf r})$ and $\hat \varphi ^\dagger ({\bf r})$ 
into $\hat n_-({\bf r})$ and $\hat n_+({\bf r})$, one obtains the
electron-positron contact density operator from equation (\ref{eq:lambda1})
or (\ref{eq:Zeff1}), in the form,
\begin{equation}\label{eq:nndr}
\int \hat n_-({\bf r})\hat n_+({\bf r})\rmd{\bf r}=
\sum _{\mu , \mu '}\sum _{\nu , \nu '}
\langle \nu '\mu '|\delta |\mu \nu \rangle \hat a^\dagger _{\nu '}
\hat a^\dagger _{\mu '}\hat a_\mu \hat a_ \nu ,
\end{equation}
where $\mu ,\mu '$ and $\nu ,\nu '$ are the electron and positron
states, respectively. The amplitude
\begin{equation}\label{eq:delta}
\langle \nu '\mu '|\delta |\mu \nu \rangle =
\int \varphi ^*_{\nu '}({\bf r}_2)\psi ^*_{\mu '}({\bf r}_1)
\delta ({\bf r}_1-{\bf r}_2) \psi _\mu ({\bf r}_1)\varphi _\nu ({\bf r}_2)
\rmd{\bf r}_1 \rmd{\bf r}_2.
\end{equation}
corresponds diagrammatically to a point-like $\delta $ vertex with two
positron and two electron (particle or hole) lines. It is related to
the annihilation amplitude (\ref{eq:mael}) by
\begin{equation}\label{eq:deldel}
\langle \nu '\mu '|\delta |\mu \nu \rangle =
\int \langle \nu '\mu '|\delta |{\bf P}\rangle \frac{\rmd^3P}{(2\pi )^3}
\langle {\bf P}|\delta |\mu \nu \rangle .
\end{equation}
Graphically, this equation is equivalent to connecting the vertices (a) or (b)
from figure~\ref{fig:ampl} with their complex conjugates (mirror images)
and ``pulling them together'' by the double-dashed lines, to form
a $\delta $ vertex with four external lines. This basic relation allows one to
relate products of terms in the diagrammatic expansion of the
annihilation amplitude $\langle f |\hat O_a({\bf P})|i\rangle $ to the
contributions to the total annihilation rate, along the lines of equations
(\ref{eq:intsum}) and (\ref{eq:lambda1}) (see below).

\subsection{Annihilation amplitudes}\label{subsec:anamp}

\begin{figure}[ht]
\begin{center}
\includegraphics*[width=15.5cm]{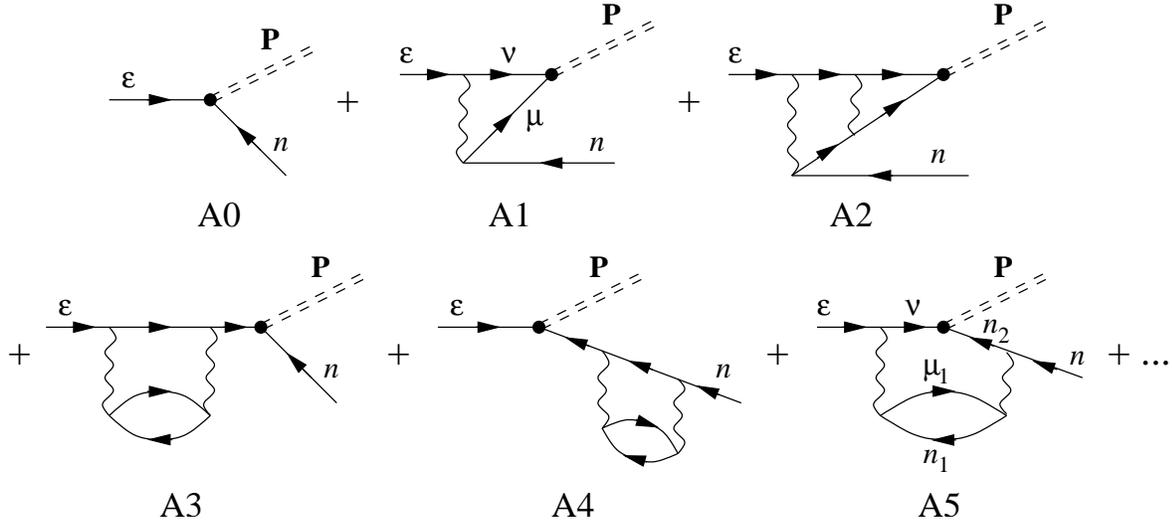}
\end{center}
\caption{Many-body theory expansion of the positron-atom annihilation
amplitude leading to the final state with a hole $n$. The line starting
with $\eps $ corresponds to the positron; $\nu $ and $\mu $ are
the intermediate positron and electron states, respectively. Wavy lines
correspond to the electron-positron or electron-electron Coulomb
interactions.}\label{fig:012}
\end{figure}

Consider a process of two-photon positron annihilation with a ground-state
atom,
\begin{equation}\label{eq:process}
e^++A\longrightarrow A^+ +2\gamma .
\end{equation}
The simplest final state $f$ of the positive ion $A^+$ is that of an atom with
a hole in an electron orbital $n$. In the lowest order the amplitude of this
process is shown by diagram (a), figure \ref{fig:ampl}. The exact amplitude
$\langle f|\hat O_a({\bf P})|i\rangle $ can be represented by a many-body
theory expansion in powers of the residual electron-positron and
electron-electron Coulomb interactions. Figure \ref{fig:012} shows the
corresponding 0th and 1st-order diagrams, together with the main types of
2nd-order diagrams.

Analytically, the first diagram in figure \ref{fig:012} is equal to
$\langle {\bf P}|\delta |n\eps \rangle $,  cf. equation (\ref{eq:mael}).
Expressions for the higher-order diagrams are constructed using standard
atomic many-body theory diagrammatic rules (see, e.g., Amusia and Cherepkov
1975), modified to account for the {\em attractive} electron-positron
Coulomb interaction. These expressions contain products of Coulomb and
$\delta $-function matrix elements divided by energy denominators. Each
energy denominator corresponds to an intermediate state separating two
interactions in the
diagram. It is given by $E-\sum \eps _{\rm part}+\sum \eps _{\rm hole}+i0$,
where the sums are over all particles\footnote{If the annihilation photons
represented by the double-dashed line, are present in the intermediate state,
their energy must also be included in $\sum \eps _{\rm part}$. Since
the electron and positron energies do not include their rest energy $mc^2$,
the energy of the photons must be given with respect to $2mc^2$.} and all
holes in the intermediate state, $E$ is the energy entering
the diagram, and the infinitesimal imaginary $i0$ defines the way of
bypassing the pole. Summation over all intermediate positron and/or electron
states is carried out. The overall sign factor for a diagram is given by
$(-1)^{N_h+N_l+N_{ep}}$, where $N_h$, $N_l$ and $N_{ep}$ are the numbers of
internal hole lines, electron-hole loops and electron-positron interactions,
respectively.

For example, the 1st-order diagram A1 in figure \ref{fig:012} corresponds to
\begin{equation}\label{eq:1st}
-\sum _{\mu , \nu }\frac{\langle {\bf P}|\delta |\mu \nu \rangle 
\langle \nu \mu |V|n\eps \rangle }{\eps -\eps _\nu
-\eps _\mu +\eps _n },
\end{equation}
where
\begin{equation}\label{eq:V}
\langle \nu \mu |V|n\eps \rangle =\int \varphi ^*_{\nu }({\bf r}_2)
\psi ^*_{\mu }({\bf r}_1) \frac{1}{|{\bf r}_1-{\bf r}_2|}\psi _n ({\bf r}_1)
\varphi _\eps ({\bf r}_2) \rmd{\bf r}_1 \rmd{\bf r}_2,
\end{equation}
is the Coulomb matrix element. Diagram A1 represent a correction to the
annihilation vertex (diagram A0) due to the electron-positron interaction.
Note that for low positron energies the denominator in equation (\ref{eq:1st})
is never zero (i.e., the intermediate state is virtual for any $\eps _\nu $
and $\eps _\mu $), hence $i0$ has been dropped.

The 2nd-order diagrams shown in figure \ref{fig:012} have a clear physical
interpretation. Thus, diagram A2 represents a higher-order correction to the
annihilation vertex, similar to diagram A1, with an extra interaction between
the annihilating electron and positron. Such contributions are very
important because the electron-positron Coulomb attraction strongly
enhances the amplitude of finding the two particles at the same point.
Diagram A3 describes polarization of the atom by the positron. This
polarization gives rise to an attractive positron-atom correlation potential
which behaves as $-\alpha _d e^2/2r^4$ at large distances, $\alpha _d$ being
the atomic dipole polarizability. It has a large effect on the incident
positron (see, e.g., Dzuba \etal 1993, 1996). Diagram A4 is a correlation
correction to the Hartree-Fock wavefunction of the hole. Given that the
Hartree-Fock approximation describes atomic ground state orbitals well, one
may expect such corrections to be relatively small. Finally, diagram A5
belongs to another type of vertex corrections in which annihilation takes
place in the presence of a virtual hole-particle excitation. We will show
below that this and similar diagrams can be neglected because of a specific
cancellation between the contributions of certain pairs of diagrams in the
photon momentum distribution.

\begin{figure}[t]
\begin{center}
\includegraphics*[width=15.5cm]{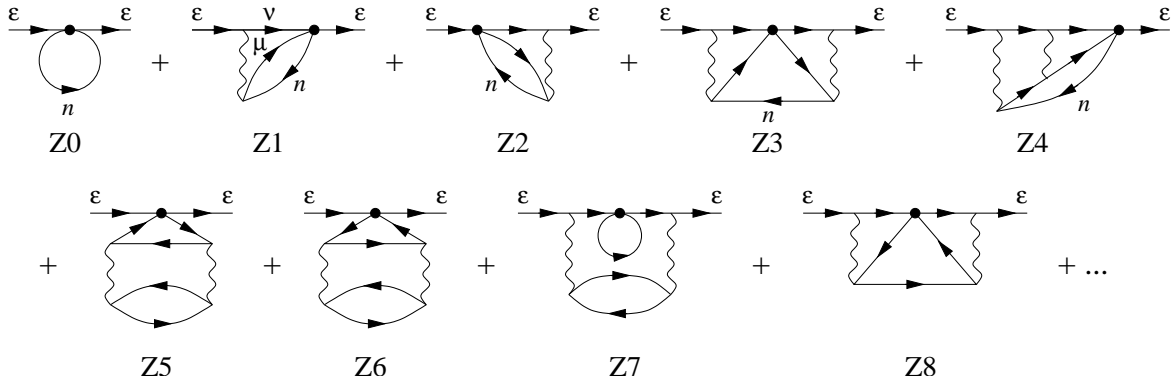}
\end{center}
\caption{Many-body theory expansion of the positron-atom annihilation rate
$Z_{\rm eff}$. Diagrams Z1--Z8 are corrections to the 0th order
annihilation vertex, diagram Z0. Corrections to the external positron lines
($\eps $) are not shown.}
\label{fig:Zeff}
\end{figure}

As explained at the end of section \ref{subsec:spra}, $Z_{\rm eff}$
can also be presented as a many-body perturbation series. If the state
$|i\rangle $ in equation (\ref{eq:Zeff1}) is that of a positron incident
on the ground state atom, $Z_{\rm eff}$ is given by the diagrammatic
expansion shown in figure~\ref{fig:Zeff}. Each of the diagrams contains a
$\delta $-function vertex corresponding to the contact density operator
(\ref{eq:nndr}) with matrix elements (\ref{eq:delta}). The 0th order diagram
Z0 describes annihilation of the positron with an electron in a Hartree-Fock
orbital occupied in the ground state. Its analytical expression,
\begin{equation}\label{eq:Zeffa}
\sum _n\langle \eps n|\delta |n\eps \rangle =\sum _n
\int |\psi _n({\bf r})|^2|\varphi _\eps ({\bf r})|^2 \rmd{\bf r},
\end{equation}
contains contributions of all occupied electron states $n$. Higher order
diagrams in figure \ref{fig:Zeff} represent correlation corrections to the
annihilation vertex. Their analytical expressions are derived using the
rules formulated above. For example, diagram Z1 corresponds to
\begin{equation}\label{eq:Zeffb}
-\sum _{\mu , \nu ,n}\frac{\langle \eps n|\delta |\mu \nu \rangle 
\langle \nu \mu |V|n\eps \rangle }{\eps -\eps _\nu
-\eps _\mu +\eps _n },
\end{equation}
cf. equation (\ref{eq:1st}). Note that in figure \ref{fig:Zeff} we do not show
corrections to the external positron lines $\eps $, like that
in diagram A3 of figure \ref{fig:012}. These corrections can be easily
included by replacing the positron state $\varphi _\eps $ in
the static field of the atom by the positron {\em Dyson orbital} which
accounts for the positron-atom correlation potential (Dzuba \etal 1996,
Gribakin and Ludlow 2004).

By analogy with equations (\ref{eq:lambda})--(\ref{eq:lambda1}) and
(\ref{eq:Zeff1}), the total annihilation rate is related to the momentum
distribution of the photon pairs by
\begin{equation}\label{eq:Zeff_O}
Z_{\rm eff}=\sum _f\int |\langle f|\hat O_a({\bf P})|i\rangle |^2
\frac{\rmd^3P}{(2\pi )^3}.
\end{equation}
This means that if the terms of the series in figure \ref{fig:012} are
multiplied by their complex conjugates and integrated over ${\bf P}$, the
expansion for $Z_{\rm eff}$ from figure \ref{fig:Zeff} must be recovered.
Indeed, it is easy to see, using equation (\ref{eq:deldel}), that the
contribution of diagram A0 to the right-hand side of equation
(\ref{eq:Zeff_O}) (i.e., A0 times its complex conjugate A0$^*$) is equal to
diagram Z0, after summation over $n$. Similarly, the product of diagrams A1
and A0$^*$,
\begin{equation}\label{eq:A0A1}
\fl
\int \langle \eps n|\delta |{\bf P}\rangle \frac{\rmd^3P}{(2\pi )^3}
\left[ -\sum _{\mu , \nu }\frac{\langle {\bf P}|\delta |\mu \nu \rangle 
\langle \nu \mu |V|n\eps \rangle }{\eps -\eps _\nu
-\eps _\mu +\eps _n }\right]
=-\sum _{\mu , \nu }\frac{\langle \eps n|\delta |\mu \nu \rangle 
\langle \nu \mu |V|n\eps \rangle }{\eps -\eps _\nu
-\eps _\mu +\eps _n },
\end{equation}
gives Z1 (after summation over $n$), while the product of A1 and A1$^*$ becomes
Z3.

However, some contributions that appear on the right-hand side of equation
(\ref{eq:Zeff_O}) do not match any diagrams of the $Z_{\rm eff}$ expansion.
In fact, their analytical expressions do not even have the
form that would identify them with a particular diagram. For example,
consider a product of diagrams A5 and A0$^*$ from figure \ref{fig:012},
\begin{equation}\label{eq:A5_A0}
\langle \eps n|\delta |{\bf P}\rangle \sum _{\nu \mu_1n_1n_2}
\frac{\langle {\bf P}|\delta |n_2\nu \rangle \langle n_1n_2|V|n\mu _1\rangle
\langle \nu \mu _1|V|n_1\eps \rangle }
{(\eps -\eps _{2\gamma} -\eps _{\mu _1}+\eps _{n_1}+\eps _{n_2})
(\eps -\eps _{\nu }-\eps _{\mu _1}+\eps _{n_1})},
\end{equation}
where $\eps _{2\gamma }=\eps +\eps _n$ is the energy of the two gamma quanta
with respect to $2mc^2$, given by the energy conservation, $E_{\gamma 1}+
E_{\gamma 2}=2mc^2+\eps +\eps _n$. It is easy to check that the
expression obtained upon integration of (\ref{eq:A5_A0}) over ${\bf P}$,
can not be drawn as a diagram consistent with the diagrammatic rules. Hence,
it also does not correspond to {\em any} diagram in the expansion of
$Z_{\rm eff}$.

\begin{figure}[ht]
\begin{center}
\includegraphics*[width=14cm]{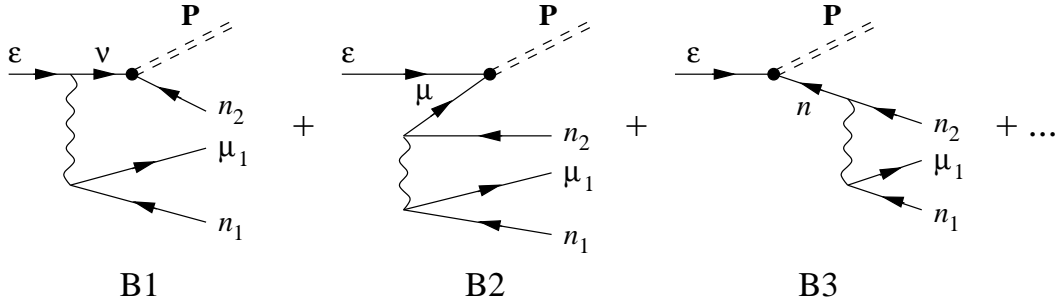}
\end{center}
\caption{Amplitude of positron annihilation leading to the final state
with two holes, $n_1$ and $n_2$, and excited electron $\mu _1$. For
$n_1\neq n_2$, exchange diagrams with the indices $n_1$ and
$n_2$ swapped and an extra factor of $-1$, must also be considered.}
\label{fig:2h1e}
\end{figure}

To solve this paradox, note that equation (\ref{eq:Zeff_O}) contains a sum
over {\em all} final states. Thus, besides the process shown in
figure~\ref{fig:012}, where positron annihilation leads to a single-hole state,
one must consider processes with other types of final states. 
Their formation is a result of extra electron-positron or electron-electron
correlations (which means that their contribution to the total annihilation
spectrum should, in general, be smaller). The simplest of such states
contains two holes and one excited electron. The corresponding amplitude in
the lowest (first) order is shown in figure \ref{fig:2h1e}.

It is easy to see how these diagrams contribute to the right-hand side of
equation~(\ref{eq:Zeff_O}). Thus, B1 times B1$^*$
(integrated over ${\bf P}$) becomes diagram Z7, B1 times the exchange
analogue of B1$^*$ results in diagram Z8, while B2 times B2$^*$
gives diagram Z5. At the same time, B1 times B3$^*$,
\begin{equation}\label{eq:B1_B3}
\sum _{\nu n}\frac{\langle n_1n_2|V|n\mu _1 \rangle 
\langle \eps n|\delta |{\bf P}\rangle
\langle {\bf P}|\delta |n_2\nu \rangle 
\langle \nu \mu_1|V|n_1\eps \rangle}
{(\eps - \eps _{2\gamma } +\eps _{n})
(\eps -\eps _\nu -\eps _{\mu _1}+\eps _{n_1})},
\end{equation}
does not lead to any $Z_{\rm eff}$ diagram after integration over ${\bf P}$.
In fact, it does not lead to any valid diagrammatic expression at all.
However, the matrix elements in equation (\ref{eq:B1_B3}) match those
in equation (\ref{eq:A5_A0}). Using the energy conservation for the process
in figure \ref{fig:2h1e},
$\eps =\eps _{2\gamma }+\eps _{\mu _1}-\eps _{n_1}-\eps _{n_2}$, we see
that the first energy denominator in (\ref{eq:B1_B3}) is equal to
$\eps _n+\eps _{\mu _1}-\eps _{n_1}-\eps _{n_2}$, while the
first energy denominator in (\ref{eq:A5_A0}) is equal to
$-\eps _n-\eps _{\mu _1}+\eps _{n_1}+\eps_{n_2}$. Hence, the two expressions
cancel term by term, as (\ref{eq:A5_A0}) is summed over the hole
states $n$, and (\ref{eq:B1_B3}) is summed over the electron and hole states
$\nu _1$, $n_1$ and $n_2$.

Note that this cancellation of the two anomalous contributions to the
total annihilation rate, $Z_{\rm eff}$, is exact. However, it occurs even
before the integration over the momentum of the photons. This means that
these contributions can be omitted when calculating the photon momentum
distribution and gamma spectrum. In particular, diagram A5 in
figure~\ref{fig:012} can be ignored. A similar cancellation also takes place
for other diagrams representing corrections to the annihilation vertex, in
which the incident positron line is connected with the final-state hole by
means of Coulomb interactions and electron-hole pairs. In all of these
diagrams annihilation occurs is the presence of one or more virtual
electron-hole pairs.

In principle, the photon spectra corresponding to different final ionic
states (e.g., those shown in figures \ref{fig:012} and \ref{fig:2h1e}) could
be considered separately. In this case, the contributions (\ref{eq:A5_A0})
and (\ref{eq:B1_B3}) are not added together, and do not cancel each other.
In practice, though, the total energy of the two photons or the final state
of the ion are usually not observed. The shift of the centre of the gamma
line due to the difference in the total photon energy is negligible (as
explained in section \ref{subsec:spra}) and the cancellation described
above does take place.


\section{Evaluation of the diagrams}\label{sec:diag}

Let us consider a positron with momentum ${\bf k}$ which collides with a
closed-shell atom in the ground state, and annihilates, creating a hole in
state $n$. The amplitude of this process is shown diagrammatically in
figure \ref{fig:012}. Let us denote it by $A_{n{\bf k}}({\bf P})$. Omitting
the QED factor $\pi r_0^2 c$ in equation (\ref{eq:wepsP}), we write the photon
energy spectrum in this process as
\begin{equation}\label{eq:wn}
w_n(\epsilon )=\frac{1}{c}\int \int _{2|\epsilon|/c}^{\infty}
|A_{n{\bf k}}({\bf P})|^2 \frac{P\rmd P\rmd\Omega _{\bf P}}{(2\pi )^3}.
\end{equation}
Integration over $\epsilon $ gives the contribution of hole $n$ to the total
annihilation rate,
\begin{equation}\label{eq:Zeffwn}
Z_{\rm eff}(n)=\int w_n(\epsilon ) \rmd \epsilon
=\int |A_{n{\bf k}}({\bf P})|^2 \frac{\rmd ^3P}{(2\pi )^3}.
\end{equation}

\subsection{Zeroth order}\label{subsec:0th}

In the 0th approximation the amplitude $A_{n{\bf k}}({\bf P})$ is given
by diagram A0, figure \ref{fig:012},
\begin{equation}\label{eq:Amp0}
A_{n{\bf k}}({\bf P})=\langle {\bf P}|\delta |n{\bf k}\rangle =
\int \rme^{-i{\bf P}\cdot {\rm r}} \psi_n({\bf r})\varphi_{{\bf k}}({\bf r}) 
\rmd {\bf r},
\end{equation}
where
\begin{equation}\label{eq:wfe}
\psi_n({\bf r})=\frac{1}{r}P_{nl}(r)Y_{lm}(\Omega),
\end{equation}
is the Hartree-Fock wavefunction of the occupied electron state in the
orbital $nl$. The positron wavefunction is normalized to a plane wave,
$\varphi_{\bf k}({\bf r})\sim e^{i{\bf k}\cdot {\bf r}}$,
and can be written as a partial-wave expansion (Landau and Lifshitz 1982),
\begin{equation}\label{eq:wfp}
\varphi_{\bf k}({\bf r})=\frac{1}{r}\sqrt{\frac{\pi}{k}}
\sum_{l_1=0}^{\infty }i^{l_1}\exp( i\delta _{l_1})
P_{\eps l_1}(r)P_{l_1}({\bf k}\cdot {\bf r}/kr) .
\end{equation} 
Here $P_{l_1}$ is the Legendre polynomial, and $P_{\eps l_1}(r)$ is the
positron radial wavefunction with energy $\eps =k^2/2$ and orbital
angular momentum $l_1$, normalized by
\begin{equation}\label{eq:wfP}
P_{\eps l_1}(r)\sim(\pi k)^{-1/2}\sin(kr-\case{1}{2}\pi l_1 + \delta_{l_1}),
\end{equation}
where $\delta _{l_1}$ is the phase shift. This normalization is to a
$\delta $-function of energy in Rydberg.

Using (\ref{eq:wfe}) and (\ref{eq:wfp}), and expanding
$\rme ^{-i{\bf P}\cdot {\bf r}}$ in spherical harmonics, we integrate over
the angular variables in equations (\ref{eq:Amp0}) and (\ref{eq:wn}), and sum
over the electronic magnetic quantum number $m$ and spin. This gives the
following expression for the spectrum of positron annihilation with an
electron from orbital $nl$,
\begin{equation}\label{eq:wnl}
w _{nl}(\epsilon)=\frac{4}{ck}\sum_{\lambda ,l_1}\int_{2|\epsilon|/c}^\infty
|A_{n\eps }^{(\lambda )}(P)|^2 P dP ,
\end{equation}
where
\begin{equation}\label{eq:redamp}
\fl A_{n\eps }^{(\lambda )}(P)=\sqrt{[\lambda ][l][l_1]}
\left(\begin{array}{ccc}\lambda & l & l_1\\
0 & 0 & 0\end{array} \right )
\int_0^{\infty} j_{\lambda}(Pr) P_{nl}(r) P_{\eps l_1}(r) \rmd r
\equiv \langle P\|\delta _\lambda \|n\eps \rangle ,
\end{equation}
is the {\em reduced} annihilation matrix element, in which
$j_\lambda (z)=\sqrt{\frac{1}{2}\pi /z}J_{\lambda +1/2}(z)$ is the spherical
Bessel function, $\lambda $ is the angular momentum
carried by the photons, and the notation $[l]\equiv 2l+1$ is used.
Equations (\ref{eq:wnl})--(\ref{eq:redamp}) are similar to equation (III.6)
of Farazdel and Cade (1977), for the angular correlation function in a
positron bound state.

\subsection{First order}\label{subsec:1st}

The 1st-order order correction A1, equation (\ref{eq:1st}), is evaluated in a
similar way, using (\ref{eq:wfp}) as the incident positron state $\eps $.
The wavefunctions of the intermediate electron and positron
states $\mu $ and $\nu $ are written as
\begin{equation}\label{eq:numu}
\psi _\mu ({\bf r})=\frac{1}{r}P_{\eps _\mu l_\mu }(r)Y_{l_\mu m_\mu }(\Omega),
\quad 
\phi _\nu ({\bf r})=\frac{1}{r}P_{\eps _\nu l_\nu }(r)Y_{l_\nu m_\nu }(\Omega),
\end{equation}
and the Coulomb interaction $V$ is also expanded in the spherical harmonics.

When the 0th and 1st-order contributions are added together, the gamma
spectrum is again given by equation (\ref{eq:wnl}), where the amplitude is now
\begin{equation}\label{eq:0+1}
A_{n\eps }^{(\lambda )}(P)=\langle P\|\delta _\lambda \|n\eps \rangle 
-\sum _{\mu ,\nu }\frac{\langle P\|\delta _\lambda \|\mu \nu \rangle
\langle \nu \mu \|V^{(\lambda )}\|n\eps \rangle }
{\eps -\eps _\nu -\eps _\mu +\eps _n}.
\end{equation}
In the above expression,
\begin{equation}\label{eq:V(lam)}
\langle \nu \mu \|V^{(\lambda )}\|n\eps \rangle =\sum _{\lambda '}
(-1)^{\lambda + \lambda'}
\left \{
\begin{array}{ccc}
\lambda  & l_1   & l     \\
\lambda' & l_\mu & l_\nu
\end{array}
\right \}
\langle \nu \mu \|V_{\lambda '}\|n\eps \rangle ,
\end{equation}
is the reduced matrix element of the Coulomb interaction within the
electron-positron pair with angular momentum $\lambda $,
\begin{equation}\label{eq:Vred}
\langle \nu \mu \|V_{\lambda '}\|n\eps \rangle =
\sqrt{[l_\nu][l_\mu ][l][l_1]}
\left (
\begin{array}{ccc}
l_\nu & \lambda' & l_1 \\
0     &    0     &  0
\end{array}
\right )
\left (
\begin{array}{ccc}
l_\mu & \lambda' & l \\
0     &    0     & 0 \\
\end{array}
\right )
R^{\lambda'}_{\nu \mu n \epsilon },
\end{equation}
is the usual reduced Coulomb matrix element, and
\begin{equation}\label{eq:Rint}
R^{\lambda'}_{\nu \mu n \epsilon }=\int \!\!\!\int P_{\eps _\nu l_\nu }(r_2)
P_{\eps _\mu l_\mu }(r_1)\frac{r_<^{\lambda '}}{r_>^{\lambda '+1}}
P_{nl}(r_1) P_{\eps l_1}(r_2)\rmd r_1 \rmd r_2,
\end{equation}
is the radial Coulomb integral. In the amplitude (\ref{eq:0+1}) the sum
over $\mu $ and $\nu $ implies summation (integration) over the orbital angular
momenta and energies of the intermediate electron and positron states.

\section{Numerical calculations and results}\label{sec:res}

In this section we present and analyse the gamma spectra obtained by using
the 0th and 1st-order terms in the amplitude, equations (\ref{eq:wnl}),
(\ref{eq:redamp}) and (\ref{eq:0+1}), for the noble gas atoms.

\subsection{Details of the calculations}\label{subsec:det}

The calculation starts by determining the electron wavefunctions of the
ground-state atom in the Hartree-Fock approximation. Subsequently, sets
of incident positron and intermediate electron and positron states in the
field of the ground-state atom are generated.

In experiments, the gamma spectra of noble gases were measured with
thermalized room-temperature positrons ($k_BT=25$ meV). At such
low energies only the positron $s$ wave contributes significantly
to the annihilation signal, higher partial waves being suppressed 
by the centrifugal barrier. The shape of the annihilation spectrum
is not sensitive to the exact value of the positron energy (as long as it
remains small), and we calculate the spectra for the positron momentum of
$k=0.05$ au, corresponding to the energy of $\eps =34$ meV
($\sim \frac{3}{2}k_BT$).

The sets of intermediate positron and electron states with orbital angular
momenta $l=0$--8 were calculated on a uniform mesh in momentum space,
starting with $k_0=0.1$ au, and increasing in steps of $\Delta k =0.2$ au,
over 59 continuous spectrum wavefunctions. The electron and positron
wavefunctions were used to compute annihilation and Coulomb matrix elements,
and construct the annihilation amplitudes (\ref{eq:0+1}) for a range of
momenta $P$. Using those, the gamma spectra for the valence and inner
subshells were calculated from equation (\ref{eq:wnl}).

It is known that single-centre expansions of the annihilation amplitudes
converge slowly with respect to the angular momenta of the electron
and positron orbitals involved (Mitroy and Ryzhikh 1998, Mitroy \etal 2002,
Gribakin and Ludlow 2002). Thus, Gribakin and Ludlow (2002) showed that the
increments in the annihilation rate upon increasing the maximal angular
momentum from $l-1$ to $l$, decrease asymptotically as
$(l+\frac12)^{-2}$. In application to the annihilation spectra, this means
that the quantity
obtained by summing over all intermediate electron and positron states with
$l$ up to $l_{\rm max}$ converges to the value for $l_{\rm max}=\infty $ as,
\begin{equation}\label{eq:fit}
w(\epsilon)^{[l_{\rm max}]}\simeq w(\epsilon)^{[\infty]}-\frac{A}
{l_{\rm max}+\frac{1}{2}},
\end{equation}
where $A$ is some constant\footnote{The 2nd term on the right-hand
side of (\ref{eq:fit}) is only the leading term, the next one being
$\propto (l_{\rm max}+\frac{1}{2})^{-2}$, and keeping $\frac 12$ next
to large $l_{\rm max}$ is simply a matter of convenience.}.
We used this relation to obtain the values of
$w(\epsilon)\equiv w(\epsilon)^{[\infty]}$ by linear extrapolation 
of $w(\epsilon)^{[l_{\rm max}]}$ from the last two points.

The total spectra are found by adding the contributions of the different
subshells. They can be compared with experimental data. Annihilation rates
obtained by integration of the spectra over $\epsilon $,
provide information on the relative contributions of various subshells to the
total $Z_{\rm eff}$, and show the importance of the correlation correction
and extrapolation over $l_{\rm max}$.

\subsection{Results for Ar}\label{subsec:Ar}

In this section we examine in detail the results of the calculations for Ar,
using it as an example. The next section will present a summary of the results
for all noble gas atoms.

\begin{figure}[!ht]
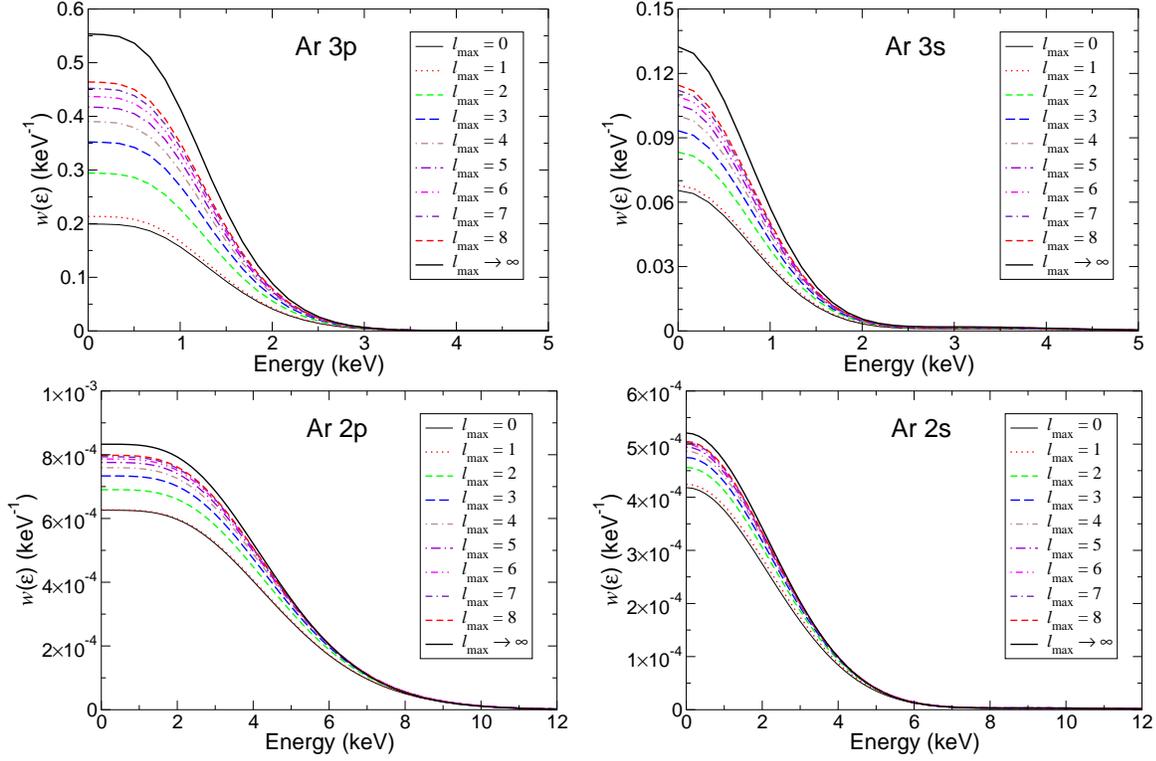

\begin{minipage}{7.5cm}
\begin{center}
\includegraphics*[height=5.cm]{Arlmax3p.eps}
\end{center}
\end{minipage}
\hspace*{0cm}
\begin{minipage}{7.5cm}
\begin{center}
\includegraphics*[height=5.cm]{Arlmax3s.eps}
\end{center}
\end{minipage}\\
\begin{minipage}{7.5cm}
\begin{center}
\includegraphics*[height=5.cm]{./Arlmax2p.eps}
\end{center}
\end{minipage}
\hspace*{0cm}
\begin{minipage}{7.5cm}
\begin{center}
\includegraphics*[height=5.cm]{./Arlmax2s.eps}
\end{center}
\end{minipage}
\caption{Gamma ray spectra $w_{nl}(\epsilon )$ for the positron annihilation
on $3p$, $3s$, $2p$ and $2s$ subshells in Ar, calculated using 0th and
1st-order diagrams. The graphs show how the result accumulates
with the increase of the maximal orbital angular momentum $l_{\rm max}$ of
the intermediate electron and positron states in the 1st-order diagram.
Thick solid curve is the result of extrapolation to
$l_{\rm max}\rightarrow \infty $, equation (\ref{eq:fit}).}
\label{fig:Arsub}
\end{figure}

Figure \ref{fig:Arsub} shows the annihilation gamma spectra for the four
outer subshells in Ar. They were obtained from equations
(\ref{eq:wnl}) and (\ref{eq:0+1}) by including the electron and positron
intermediate states with orbital angular momenta up to $l_{\rm max}$
in the 1st-order term, with $l_{\rm max}=0$--8. Since the spectra are
symmetric, $w(-\epsilon )=w(\epsilon )$, only the positive energies
are shown. The $1s$ subshell is not included, as its contribution is
small and does not affect the results for the total spectrum. 

\begin{figure}[ht]
\begin{center}
\includegraphics[width=9cm]{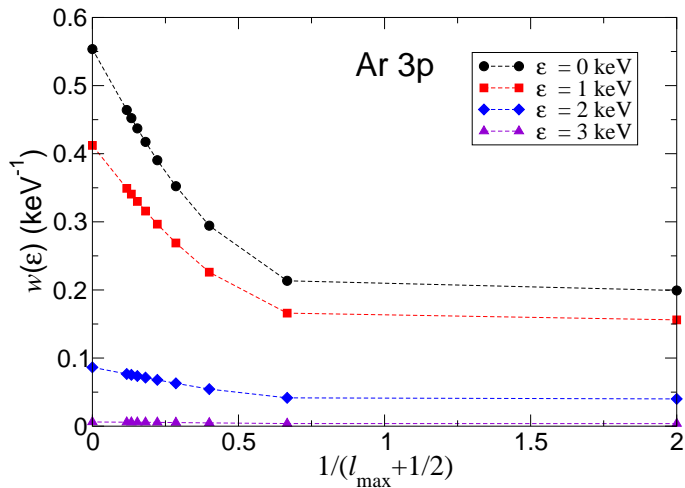}
\end{center}
\caption{Extrapolation of the spectral density $w_{3p}(\epsilon)$ with respect
to $l_{\rm max}$. Different symbols show the spectra calculated for
$l_{\rm max}=0$--8 and $l_{\rm max}\rightarrow \infty $, for different
energies: full circles, $\epsilon = 0$ keV; squares, $\epsilon = 1$ keV;
diamonds, $\epsilon = 2$ keV; triangles,  $\epsilon = 3$ keV. The values
at $(l_{\rm max}+\frac12 )^{-1}=0$ are obtained by linear extrapolation,
equation (\ref{eq:fit}), from the values for $l_{\rm max}=7$ and 8.}
\label{fig:extra}
\end{figure}

In the plots the lowest (thin solid) curve labelled $l_{\rm max}=0$
corresponds to the 0th-order result (for $3p$ and $2p$), or is very close
to it (for $3s$ and $2s$), since the contribution of the intermediate states
with the zero angular momentum is very small. For greater angular momenta
the contributions of successive $l$ decrease slowly, as expected.
The spectra obtained by extrapolation to $l_{\rm max}\rightarrow \infty $
(thick solid curves) are noticeably larger than those obtained with
$l_{\rm max}=8$. For the two outer subshells, extrapolation increases the
spectra by about 20\% at small $\epsilon $. Details of the extrapolation
procedure are illustrated by figure \ref{fig:extra}. In agreement with
equation (\ref{eq:fit}), the dependence of $w^{[l_{\rm max}]}(\epsilon )$ on
$(l_{\rm max}+\frac{1}{2})^{-1}$ is close to linear at large $l_{\rm max}$.

Figure \ref{fig:Arsub} shows that the effect of electron-positron
correlations described by the 1st-order correction, is strongest for the
$3p$ and $3s$ subshells which have small ionization potentials
($I_{3p}=15.76$ eV, $I_{3s}=29.24$ eV). The electrons in the inner shells are
much more strongly bound ($I_{2p}=249$ eV, $I_{3s}=326$ eV). This makes it
harder for the positron to perturb their motion, and reduces the effect of
correlations.
Nevertheless, the 1st-order correction leads to a noticeable increase of
the annihilation signal for them as well. The larger binding energies also
mean that $2s$ and $2p$ electrons move at greater speeds and have a broader
momentum distribution. As a result, their gamma spectra are wider, as seen
in figure \ref{fig:Arsub}. Because of the repulsion from the nucleus, the
positron wavefunction has a weaker overlap with the inner-shell electron
orbitals, and their contribution to the positron-atom annihilation rate is
much smaller than that of the valence shell.

Tables \ref{tab:Arfwhm} and \ref{tab:Arzeff} quantify the effect of the
1st-order correlation correction on the shapes of the gamma
spectra and annihilation rates. Table \ref{tab:Arfwhm} gives the full widths
at half-maximum (FWHM) of the annihilation spectra for different orbitals.
It shows that the 1st-order correction reduces the FWHM values of the spectra.
This can be explained by noticing that in the 1st-order diagram (A1) the
positron annihilates with an excited electron ``pulled out'' of the atom.
The wavefunction of such excited electron is less localized than the bound
state wavefunction. The corresponding typical electron momenta are lower,
making for smaller Doppler shifts and a narrower gamma spectrum.

\begin{table}[!ht]
\caption{Full widths at half-maximum of the gamma spectra for Ar (in keV).}
\label{tab:Arfwhm}
\begin{indented}
\item[]\begin{tabular}{@{}llll}
\br
&&\centre{2}{0th + 1st order}\\
\ns
&&\crule{2}\\
Subshell & 0th order & $l_{max}=8$ & $l_{max} \rightarrow \infty$ \\
\mr
$2s$ & 5.18    & 5.20 & 5.15\\
$2p$ & 9.42    & 9.29 & 9.18\\
$3s$ & 1.86    & 1.84 & 1.82\\
$3p$ & 2.89    & 2.75 & 2.71\\
\mr
Total& 2.65    & 2.58 & 2.55\\
\br
\endTable

\begin{table}
\caption{Contribution of the outer and inner-shell electrons to
$Z_{\rm eff}$ in Ar.}
\label{tab:Arzeff}
\begin{indented}
\item[]\begin{tabular}{@{}llll}
\br
& &\centre{2}{0th + 1st order}\\
\ns
&&\crule{2}\\
Subshell & 0th order & $l_{max}=8$&$l_{max} \rightarrow \infty$\\
\mr
$2s$ & 0.00237 & 0.00284 & 0.00289\\
$2p$ & 0.00621 & 0.00777 & 0.00801\\
$3s$ & 0.13585 & 0.23265 & 0.26610\\
$3p$ & 0.60562 & 1.34068 & 1.57887\\
\mr
Total& 0.75005 & 1.58394 & 1.85587\\
\br
\endTable

The results shown in table \ref{tab:Arzeff} are the breakdown of the
annihilation rate $Z_{\rm eff}$ for the $2s, 2p, 3s$ and $3p$ subshells.
Partial $Z_{\rm eff}$ were obtained by integration of the corresponding
spectral densities $w_{nl}(\epsilon )$, equation (\ref{eq:Zeffwn}). According
to the table, extrapolation over $l_{\rm max}$ beyond $l_{\rm max}=8$
increases the $Z_{\rm eff}$ values by about 15\% for the $3s$ and $3p$
orbitals, and by 2--3\% for the $n=2$ orbitals. The final values show that
adding the correlation correction more than doubles the $Z_{\rm eff}$ value
for the $n=3$ shell, compared with the 0th-order calculation. It also
increases the contributions of the $2s$ and $2p$ subshells, by 22\% and 28\%,
respectively. Given their large binding energies, this is a remarkably strong
correlation effect. In particular, it is much greater than their contribution
of the $n=2$ electrons to the positron-atom correlation potential, which can
be estimated from their contribution to the dipole polarizability of Ar.

The size of the correlation effect for the inner-shell electrons can be
compared with the {\em enhancement factor} for positron annihilation on
hydrogen-like positive ions (Novikov \etal 2004). This factor
is defined as the ratio of $Z_{\rm eff}$ calculated with an accurate
correlated wavefunction to the value obtained using a product of electron and
positron densities $n_\pm ({\bf r})$:
$Z_{\rm eff}=\int n_-({\bf r})n_+({\bf r})\rmd {\bf r}$. The latter equation
would be correct if the total wavefunction of the system was an uncorrelated
product of the electron and positron wavefunctions, cf. equation
(\ref{eq:Zeff1}). In the many-body theory approach, the enhancement factor
can be defined as the ratio of $Z_{\rm eff}$ calculated from the expansion
in figure \ref{fig:Zeff} to the value obtained from the 0th-order diagram Z0,
equation (\ref{eq:Zeffa}) (Dzuba \etal 1996, Gribakin and Ludlow 2004).
This enhancement factor describes the effect of short-range electron-positron
correlations (Dzuba \etal 1996). For the annihilation of $s$-wave positrons
on the valence electrons of noble gas atoms and hydrogen, it ranges from
about 2.5 for Ne to 6 for Xe and H (Dzuba \etal 1996, Gribakin and
Ludlow 2004, Ludlow 2003). According to Novikov \etal (2004), the enhancement
factors for B$^{4+}$ and F$^{8+}$ are 1.41 and 1.20, while the
electron ionization potentials in these systems are 340 and 1100 eV,
respectively. The ionization potentials of the $2s$ and $2p$ subshells of Ar
are similar to the first of these values, and the enhancement factors for
$Z_{\rm eff}$ due to the 1st-order correlation correction in table
\ref{tab:Arfwhm} are of comparable magnitude, 1.22 ($2s$) and 1.28 ($2p$).

In spite of the large increase of $Z_{\rm eff}$ above the 0th-order result,
the final total in table \ref{tab:Arzeff} (1.86) is still much smaller
than the experimental $Z_{\rm eff}=26.77\pm 0.09$ obtained with room
temperature positrons (Coleman \etal 1975). Of course, the large size of the
1st-order diagram means that higher-order corrections are also important,
especially for the valence and subvalence orbitals. The main contribution
here comes from electron-positron ladder diagrams (e.g., A2 in figure
\ref{fig:012}). Another important effect is the distortion of the positron
wavefunction by the positron-atom correlation potential (such as that shown
by diagram A3). As mentioned in section \ref{subsec:anamp}, this strong
nonperturbative effect can be accounted for by calculating the positron
wavefunction from the Dyson equation. All-order summation
of the ladder diagram series and the Dyson equation for the positron have
been incorporated in the calculations of $Z_{\rm eff}$ (Gribakin and Ludlow
2004, Ludlow 2003), yielding good agreement with accurate calculations for
hydrogen and experiment for the noble gases. We are currently applying a
similar approach to the calculation of gamma spectra (Dunlop and
Gribakin 2005).

Figure \ref{fig:Arshpe} shows our final gamma spectra for the
$2s$, $2p$, $3s$ and $3p$ subshells. Addition of the partial spectra leads to a
smoother, almost featureless total. The contributions of the inner shells are
very small near the centre of the line. However, they noticeably push 
the wings up at $|\epsilon |>4$ keV, and contribute to a overall
non-Gaussian appearance of the spectrum. A similar effect of
{\em core annihilation} is a common feature in ACAR and Doppler broadening
spectra of positron annihilation in solids (see, e.g., Lynn \etal 1977,
Alatalo \etal 1995, 1996, Asoka-Kumar \etal 1996, Eshed \etal 2002).

\begin{figure}[ht]
\begin{center}
\includegraphics[width=9cm]{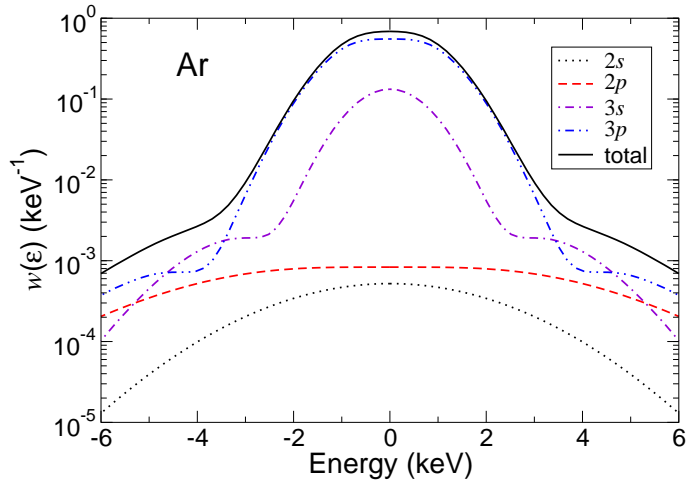}
\end{center}
\caption{Gamma spectra for $2s$, $2p$, $3s$ and $3p$ subshells in Ar obtained
in the 0th + 1st-order calculation: \dotted , $2s$; \broken , $2p$;
\chain , $3s$; \dashddot , $3p$; \full , total.}
\label{fig:Arshpe}
\end{figure}

Let us now compare the total gamma spectrum with experiment. In Iwata 1997b,
parameters of a large number of gamma spectra for atomic and molecular species
were reported. The shapes of the spectra were determined by fitting the
observed spectra with a two-Gaussian function,
\begin{equation}
\label{eq:exp}
q(E)=\exp \left [-\left(\frac{E-E_0}{a\Delta E_1}\right)^2\right]
+D \exp \left [-\left(\frac{E-E_0}{a\Delta E_2}\right)^2\right],
\end{equation}
convolved with the detector response function (see also Iwata 1997c). In
equation (\ref{eq:exp}), $a=(4 \ln 2)^{-1/2}$, and the fitting parameters
are the line centre $E_0$, the FWHM of the two Gaussians, $\Delta E_1$ and
$\Delta E_2$, and the relative weight of the 2nd Gaussian $D$.
The line-shape parameters for the noble gas atoms are given
in table \ref{tab:gausval}.

\begin{table}[ht]
\caption{$\gamma$-ray line-shape parameters from two-Gaussian fits
(Iwata \etal 1997b,c).}\label{tab:gausval}
\begin{indented}
\item[]\begin{tabular}{@{}llll}
\br
Atom & $\Delta E_1$ (keV) & $\Delta E_2$ (keV) & $D$ \\
\mr
He & 2.15 & 3.90 & 0.177\\
Ne & 3.14 & 6.12 & 0.060\\
Ar & 2.25 & 7.27 & 0.010\\
Kr & 2.02 & 6.86 & 0.016\\
Xe & 1.80 & 5.03 & 0.033\\
\br
\end{tabular}
\end{indented}
\end{table}

In figure \ref{fig:Arexp} we compare the experimental spectrum for Ar in the
form (\ref{eq:exp}), with the total spectra obtained in the 0th and
0th+1st-order approximations. For the purpose of comparison, all spectra are
normalized to unity at zero energy shift $\epsilon $. As discussed above,
inclusion of the 1st-order correction makes the spectrum narrower. It is also
clear that the enhancement due to this contribution is stronger near the
centre of the line and weaker in the wings. As a result, the wings of the
0th+1st-order spectrum in figure \ref{fig:Arexp} appear to be suppressed.
The experimental data clearly favour the curve which includes the
correlation correction.

\begin{figure}[ht]
\begin{center}
\includegraphics[width=9cm]{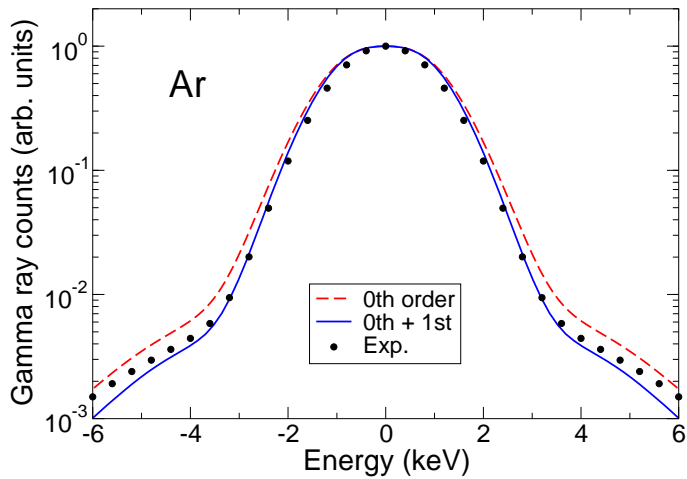}
\end{center}
\caption{Comparison of the total spectrum for Ar with experimental data. All
spectra have been normalised to unity at $\epsilon =0$. Theory:
\broken , 0th order; \full , 0th+1st order. Experiment:
\fullcircle , Iwata \etal 1997b,c.}\label{fig:Arexp}
\end{figure}

It must be said though, that the overall {\em shape} of the gamma spectrum
and, in particular, its FWHM value, appear to be very robust. They are much
less sensitive to the correlation correction than the absolute magnitude of
the annihilation rate (as seen earlier in tables \ref{tab:Arfwhm} and
\ref{tab:Arzeff}). The reason for such robustness of the shapes of
the gamma spectra is that for slow positrons the momenta $\bf {P}$ of the
annihilating electron-positron pair are determined mainly by the momentum
distribution of the bound electron. Such distribution is described well even
at the Hartree-Fock level. This explains why Iwata \etal (1997a) were able to
obtain good fits of the experimental data with crude 0th-order spectra,
by adjusting the relative amounts of the outer and inner-shell annihilation.

\subsection{Summary of results for all noble gas atoms}

Table \ref{tab:allfwhm} shows the FWHM values of the gamma spectra calculated
using the 0th and 0th+1st-order approximations, as described in sections
\ref{subsec:det} and \ref{subsec:Ar}. For all atoms (except He) the
spectrum contains the contributions of the two outermost shells, e.g.,
for Xe, annihilation on $4s$, $4p$, $4d$, $5s$ and $5p$ electrons has been
considered. For heavier noble gas atoms the contribution of
inner-shell annihilation becomes more important (Iwata 1997a). However, it
never exceeds few per cent, and has only a small effect on the FWHM values
shown in table \ref{tab:allfwhm}. We see that for all atoms except He, the
addition of the correlation correction improves the agreement with experiment.

\begin{table}[ht]
\caption{FWHM of the gamma spectra for the noble gas atoms (in keV).}
\label{tab:allfwhm}
\begin{indented}
\item[] \begin{tabular}{@{}llll}
\br
&\centre{2}{Theory}&\\
\ns
&\crule{2}\\
Atom  & 0th order & 0th+1st & \centre{1}{Experiment$^{\rm a}$}\\
\mr
He   &2.53 & 2.35 & $2.50\pm 0.03$\\
Ne   &3.82 & 3.63 & $3.36\pm 0.02$\\
Ar   &2.65 & 2.55 & $2.30\pm 0.02$\\
Kr   &2.38 & 2.30 & $2.09\pm 0.02$\\
Xe   &2.06 & 1.99 & $1.92\pm 0.02$\\
\br
\end{tabular}
\item[]$^{\rm a}$ Values obtained from Gaussian fits to the data
(Van Reeth \etal 1996, Iwata \etal 1997b).
\end{indented}
\end{table}

The changes in $Z_{\rm eff}$ (table \ref{tab:allzeff}) due to the 1st-order
correction to the annihilation amplitude, are more pronounced.
The corresponding enhancement of the annihilation rate ranges from a factor
of 1.9 in Ne to 2.9 in Xe. It is larger for heavier noble gas atoms, where
the electrons are less strongly bound, making the correlation effect greater.
Nevertheless, as in Ar, the final $Z_{\rm eff}$ values fall far short of
experiment. Higher-order corrections to the annihilation vertex ultimately
increase it above the 0th-order result by factors of 2.5 to 6, with a further
increase coming from the distortion of the positron wavefunction by the
positron-atom correlation potential (Dzuba \etal 1996, Ludlow 2003). The
latter effect becomes progressively stronger in Ar, Kr and Xe, due to the
existence of virtual positron states with energies closer to zero
(Dzuba \etal 1996).

\begin{table}[ht]
\caption{$Z_{\rm eff}$ values for the noble gas atoms.}
\label{tab:allzeff}
\begin{indented}
\item[]\begin{tabular}{@{}llll}
\br
&\centre{2}{Theory}&\\
\ns
&\crule{2} &\\
Atom&0th &0th+1st & Experiment\\
\mr
He & 0.688 & 1.395 & $3.94\pm 0.02$$^{\rm a}$\\
Ne & 0.975 & 1.850 & $5.99\pm 0.08$$^{\rm a}$\\
Ar & 0.750 & 1.856 & $26.77\pm 0.09$$^{\rm a}$\\
Kr & 0.703 & 1.870 & $64.6\pm 0.08$$^{\rm a}$, $65.7\pm 0.3$$^{\rm b}$\\
Xe & 0.645 & 1.887 & 400--450$^{\rm b}$, $401\pm 20$$^{\rm c}$ \\
\br
\end{tabular}
\item[] $^{\rm a}$ Measured in the gas (Coleman \etal 1975).
\item[] $^{\rm b}$ Measured in the gas, with small amounts of He or H$_2$
added to Xe to improve positron thermalization (Wright \etal 1985)
\item[] $^{\rm c}$ Measured in the positron trap (Murphy and Surko 1990).
\end{indented}
\end{table}

Finally, in figure \ref{fig:allexp} the gamma spectra calculated in the 0th
and 0th+1st-order approximations are compared with the two-Gaussian fits
of the experimental data (Iwata 1997b,c). In all cases the inclusion of
the vertex correction leads to better agreement with experiment. In
particular, the relative contribution of the ``shoulders'' associated
with the inner-shell contributions in Kr and Xe, is described more accurately.
As in Ar, the enhancement of the annihilation rates due to correlations is
greater for the outer $np$ and $ns$ electrons than for the inner
$(n-1)d$, $(n-1)p$ and $(n-1)s$ subshells. When the total spectra are
normalized to unity at $\epsilon =0$, this results in suppression of the
wings.

\begin{figure}[!ht]
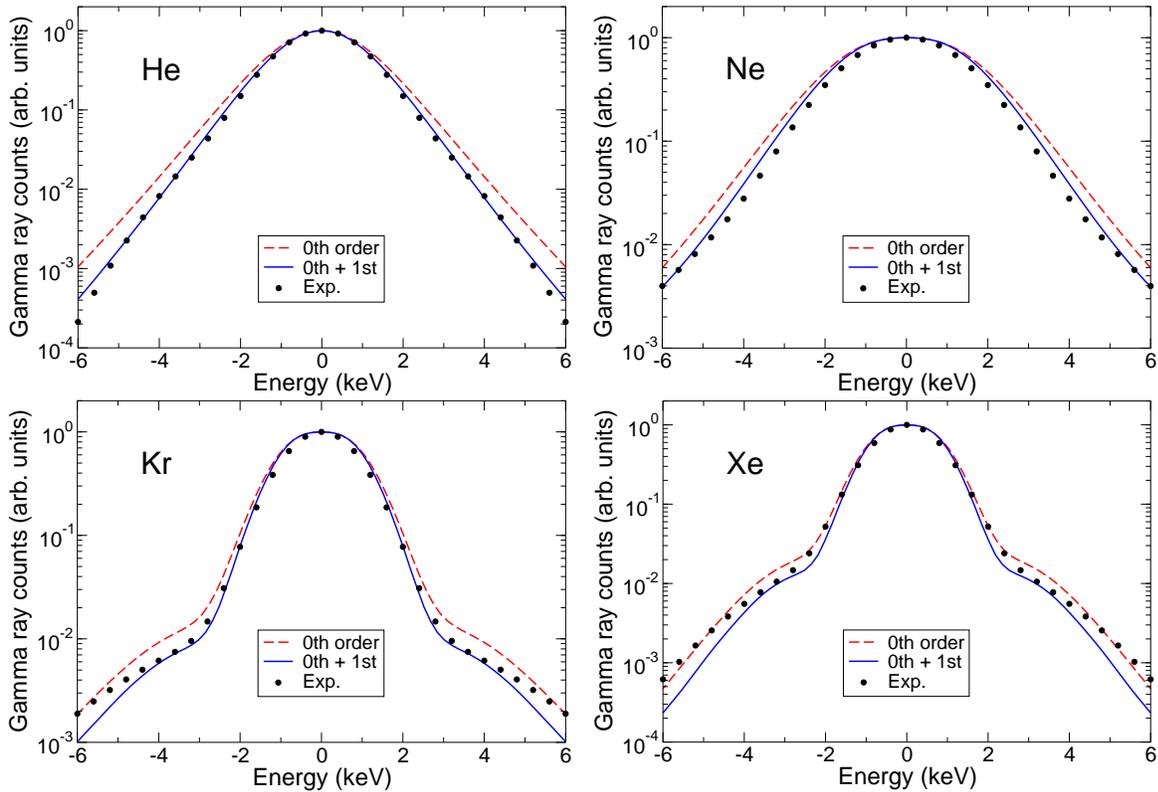

\begin{minipage}[h]{7.5cm}
\begin{center}
\includegraphics*[width=7.5cm]{Helmaxl.eps}
\end{center}
\end{minipage}
\hspace*{0cm}
\begin{minipage}[h]{7.5cm}
\begin{center}
\includegraphics*[width=7.5cm]{Nelmaxs.eps}
\end{center}
\end{minipage}\\
\begin{minipage}[h]{7.5cm}
\begin{center}
\includegraphics*[width=7.5cm]{Krlmax.eps}
\end{center}
\end{minipage}
\hspace*{0cm}
\begin{minipage}[h]{7.5cm}
\begin{center}
\includegraphics*[width=7.5cm]{Xelmax.eps}
\end{center}
\end{minipage}
\caption{\label{allexp}Comparison of the total gamma spectra for He, Ne, Kr
and Xe with experiment. The spectra have been normalized to unity at
$\epsilon =0$. Theory: \broken , 0th order; \full , 0th+1st order.
Experiment: \fullcircle , Iwata \etal 1997b,c.}
\label{fig:allexp}
\end{figure}

\section{Summary and outlook}

In this work we have shown how a many-body perturbation series can be
developed for the amplitude of positron-atom annihilation into photons with
the total momentum ${\bf P}$. We have also shown how this series converts into
a diagrammatic expansion of the annihilation rate $Z_{\rm eff}$, upon
integration over ${\bf P}$ and summation over the final states.

Expressions for the 0th and 1st-order diagrams in the annihilation
amplitude have been derived, in terms of reduced $\delta $-function and
Coulomb matrix elements. These contributions have been evaluated numerically
for the valence and inner subshells of the noble gas atoms. We have
demonstrated the importance of extrapolation over the angular momentum of
the intermediate electron and positron states. A comparison of the
calculated and measured gamma spectra has confirmed that inclusion of
the correlation correction improves agreement with experiment.
The correlation correction also leads to a sizeable enhancement of the
annihilation rates, especially for the valence and subvalence orbitals.

The large role played by the 1st-order diagram means that higher order
contributions must be considered. Thus, one needs to include
vertex corrections containing the electron-positron ladder diagrams
to all orders. One also needs to improve the incident positron wavefunction
by taking into account the positron-atom correlation potential. So far, this
programme has been realised in the calculations of positron-atom annihilation
(Gribakin and Ludlow 2004, Ludlow 2003). Work is under way to implement this
for the gamma spectra. This will allow us to obtain accurate gamma spectra
for positron-atom annihilation. In particular, a much more accurate
determination of the annihilation fraction of the inner-shell
electrons in the noble gas atoms will be possible.

It is expected that our many-body theory work will provide further insights
into the physical mechanisms which determine the shapes and intensities 
of the gamma spectra. We hope that this information, and possibly even some
of the methods we are developing, will be useful for the studies of positron
annihilation in more complex systems, such as molecules, clusters and
condensed matter.

\section*{Acknowledgements}

GG is grateful to J. Ludlow, A. V. Korol, M. G. Kozlov and M. Ya. Amusia
for the useful discussions concering the many-body theory expansion of
the annihilation amplitude. LJMD would like to thank DEL Northern Ireland
for support in the form of a PhD studentship.

\appendix

\section*{Appendix: Annihilation amplitude in the coordinate representation}
\setcounter{section}{1}

Suppose the initial state of the system is that of a positron with momentum
${\bf k}$ incident on a ground-state atom. The amplitude of finding the
positron and one of the electrons at the same point and having a total
momentum ${\bf P}$ in an annihilation event which leaves the ion in the
final state $f$, is
\begin{eqnarray}\label{eq:Af}
A_{f {\bf k}}({\bf P})&=\frac{1}{\sqrt{N}}
\sum_{i=1}^N(-1)^i\int \Psi^{*}_f({\bf r}_1,\ldots,{\bf r}_{i-1},
{\bf r}_{i+1},\ldots,{\bf r}_N) \rme^{-i{\bf P}\cdot ({\bf r}+{\bf r}_i)/2}
\nonumber \\
&\times \delta({{\bf r}-{\bf r}_i})
\Psi_{{\bf k}}({\bf r}_1,\ldots,{\bf r}_N,{\bf r})
\rmd{\bf r}_1 \dots \rmd{\bf r}_N \rmd{\bf r}.
\end{eqnarray}
Here $\Psi_{\bf k}({\bf r}_1, {\bf r}_2,\ldots,{\bf r}_N,{\bf r})$ is the
initial wavefunction of positron coordinate ${\bf r}$ and $N$ electron
coordinates ${\bf r}_i$, and $\Psi_f({\bf r}_1, \ldots, {\bf r}_{N-1})$
is the wavefunction of the ion in the final state $f$. The initial
state is normalized at large positron-atom separations according to
$\Psi_{\bf k}({\bf r}_1, \ldots,{\bf r}_N, {\bf r} )\simeq
\Phi _0({\bf r}_1, \ldots, {\bf r}_N)\rme^{i{\bf k\cdot r}}$, where $\Phi_0 $
is the target ground state wavefunction. Since the positron can annihilate
with any target electron, the terms in the sum over $i$ are antisymmetrized
by the $(-1)^i$ factor. Formula (\ref{eq:Af}) is analogous to equation (2)
of Mitroy \etal (2002).

To obtain the total annihilation rate, the amplitude (\ref{eq:Af}) must be
squared and integrated over ${\bf P}$, and a sum over all final states of
the ion should be taken:
\begin{eqnarray}
Z_{\rm eff}&=\sum_f\int |A_{f{\bf k}}({\bf P})|^2 \frac{\rmd^3 {\bf P}}
{(2 \pi)^3} \nonumber \\
 &=\int \sum _{i=1}^{N}\delta ({\bf r}-{\bf r}_i) 
\left | \Psi_{\bf k} ({\bf r}_1, \dots ,{\bf r}_n,{\bf r}) \right|^2
\rmd{\bf r}_1, \dots ,\rmd{\bf r}_n, \rmd{\bf r}.
\end{eqnarray}

In the lowest approximation the initial-state wavefunction can be written as a
product of the positron and target wave-functions, 
\begin{equation}
\Psi_{\bf k}({\bf r}_1,\ldots,{\bf r}_N,{\bf r})=
\Phi_0({\bf r}_1,\ldots,{\bf r}_N)\varphi_{\bf k}({\bf r})
\end{equation}
where $\Phi_0$ is the wavefunction of the target in the Hartree-Fock
approximation (Slater determinant) and $\varphi_{\bf k}({\bf r})$ is the
incident positron wavefunction.  Assuming that $\Psi_f$ is also a Slater 
determinant in which a single-particle electron state $n$ is missing,
we obtain the annihilation amplitude
\begin{equation}
\label{zeroamp}
A_{n {\bf k}}({\bf P})=\int \rme^{-i{\bf P}\cdot {\bf r}} \psi_{n}({\bf r})
\varphi_{{\bf k}}({\bf r}) \rmd {\bf r},
\end{equation}
which is identical to the 0th-order expression (\ref{eq:Amp0}).

\References
\item[]Alatalo M, Kauppinen H, Saarinen A, Puska M J, M\"akinen J,
Hautoj\"arvi P and Nieminen R M 1995 \PR{\it B} {\bf 51} 4176
\item[]Alatalo M, Barbiellini B, Hakala M, Kauppinen H, Korhonen T, Puska M J,
Saarinen A, Hautoj\"arvi P and Nieminen R M 1996 \PR{\it B} {\bf 54} 2397
\item[]Amusia M Ya and Cherepkov N A 1975 {\it Case Studies in Atomic Physics}
{\bf 5} 47
\item[]Amusia M Ya, Cherepkov N A, Chernysheva L V and Shapiro S G 1976
\JPB {\bf 9} L531
\item[]Arponen  J and Pajanne E 1979 {\it J. Phys. F: Metal. Phys.} {\bf 9}
2359
\item[]Asoka-Kumar P, Alatalo M, Ghosh V J, Kruseman A C, Nielsen Band Lynn K G
1996 \PRL {\bf 77} 2097.
\item[]Berestetskii V B, Lifshitz E M and Pitaevskii L P 1982
{\em Quantum Electrodynamics} (Oxford: Pergamon)
\item[]Carbotte J P 1967 \PR {\bf 155} 197 
\item[]Chang Lee 1957 {\it Zh. Eksp. Teor. Fiz.} {\bf 33} 365
(Chang Lee 1958 {\it Sov. Phys.--JETP} {\bf 6} 281)
\item[]Coleman P G, Griffith T C, Heyland R G and Killeen T L 1975 \JPB
{\bf 8} 1734
\item[]Coleman P G, Rayner S, Jacobsen F M, Charlton M and West R N 1994
\jpb {\bf 27} 981
\item[]Dunlop L J M and Gribakin G F 2005 \PR{\it A} to be submitted
\item[]Dzuba V A, Flambaum V V, King W A, Miller B N and Sushkov O P 1993
{\it Phys. Scr.} {\bf T46} 248
\item[]Dzuba V A, Flambaum V V, Gribakin G F and King W A 1995 \PR{\it A}
{\bf 52} 4541
\item[]Dzuba V A, Flambaum V V, Gribakin G F and King W A 1996 \jpb {\bf 29}
3151
\item[]Eshed A, Goktepeli S, Koymen A R, Kim S, Chen W C, O'Kelly D J,
Sterne P A and Weiss A H 2002 \PRL {\bf 89} 075503
\item[]Farazdel A and Cade P E 1977 {\it J. Chem. Phys.} {\bf 66} 2612 
\item[]Ferrel R A 1956 \RMP {\bf 28} 308
\item[]Fraser P A 1968 {\it Adv. At. Mol. Phys} {\bf 4} 63
\item[]Gribakin G F and King W A 1996 {\it Can. J. Phys.} {\bf 74} 449
\item[]Gribakin G F and Ludlow J 2002 \jpb {\bf 35} 339
\item[]Gribakin G F and Ludlow J 2004 \PR{\it A} {\bf 70} 032720 
\item[]Iwata K, Gribakin G F, Greaves R G and Surko C M 1997a \PRL
{\bf 79} 39
\item[]Iwata K, Greaves R G and Surko C M 1997b \PR{\it A} {\bf 55} 3586
\item[]Iwata K 1997c {\it Phd. Thesis} (San Diego, Univeristy of California)
\item[]Johnson W R, Buss D J, and Carroll C O 1964 \PR {\bf 135} A1232  
\item[]Kahana S 1963 \PR {\bf 129} 1622
\item[]Landau L D and Lifshitz E M 1977 {\it Quantum Mechanics}
(Oxford: Pergamon)
\item[]Ludlow J 2003 {\it Phd Thesis} (Belfast: Queen's University)
\item[]Lynn K G, MacDonald J R, Boie R A, Feldman L C, Gabbe J D, Robbins M F,
Bonderup E and Golovchenko J 1977 \PRL {\bf 38} 241
\item[]McEachran R P, Stauffer A D and Campbell L E M 1980 \JPB {\bf 13} 1281
\item[]Mitroy J and Ryzhikh G G 1998 \jpb {\bf 32} 2831
\item[]Mitroy J, Bromley M W J and Ryzhikh G G 2002 \jpb {\bf 35} R81
\item[]Murphy T J and Surko C M 1990 \jpb {\bf 23} L727
\item[]Novikov S A, Bromley M W J and Mitroy J 2004 \PR{\it A} {\bf 69} 052702 
\item[]Puska M J and Nieminen R M 1994 \RMP {\bf 66} 841
\item[]Surko C M, Gribakin G F and Buckman S J 2005 \jpb {\bf 38} R57
\item[]Tang S, Tinkle M D, Greaves R G and Surko C M 1992 PRL {\bf 68} 3793
\item[]Van Reeth P, Humberston J W, Iwata K and Surko C M 1996 \jpb {\bf 29}
L465
\item[]Wright G L, Charlton M, Griffith T C and Heyland G R 1985 \jpb {\bf 18}
4327
\endrefs

\end{document}